\documentclass[%
 reprint,
 amsmath,amssymb,
 aps,
 superscriptaddress,
 prl,
]{revtex4-2}

\usepackage{graphicx}% Include figure files
\usepackage{dcolumn}% Align table columns on decimal point
\usepackage{bm}% bold math
\usepackage{hyperref}% add hypertext capabilities
\hypersetup{colorlinks, allcolors=blue}
\usepackage{xcolor}
\usepackage{physics}
\usepackage{siunitx}
\DeclareSIUnit\rydberg{Ry}
\DeclareSIUnit\angstrom{\text{Å}}
\usepackage{tablefootnote}
\usepackage{threeparttable}
\begin{document}

\preprint{APS/123-QED}

\title{One-Dimensional Spin-Polarised Surface States - A Comparison of Bi(112) with Other Vicinal Bismuth Surfaces}

\author{Anna Cecilie Åsland}
\author{Johannes Bakkelund}
\affiliation{Department of Physics, Norwegian University of Science and Technology (NTNU), NO-7491 Trondheim, Norway.}

\author{Even Thingstad}
\affiliation{Department of Physics, Norwegian University of Science and Technology (NTNU), NO-7491 Trondheim, Norway.}
\affiliation{Department of Physics, University of Basel, Klingelbergstrasse 82, CH-4056 Basel, Switzerland.}

\author{Håkon I. Røst}
\affiliation{Department of Physics, Norwegian University of Science and Technology (NTNU), NO-7491 Trondheim, Norway.}

\author{Simon P. Cooil}
\affiliation{Centre for Materials Science and Nanotechnology, University of Oslo, Oslo 0318, Norway.}

\author{Jinbang Hu}
\affiliation{Department of Physics, Norwegian University of Science and Technology (NTNU), NO-7491 Trondheim, Norway.}

\author{Ivana Vobornik}
\author{Jun Fujii}
\affiliation{Istituto Officina dei Materiali (IOM)-CNR, Area Science Park-Basovizza, S.S. 14 Km 163.5, 34149 Trieste, Italy.}

\author{Asle Sudbø}
\affiliation{Department of Physics, Norwegian University of Science and Technology (NTNU), NO-7491 Trondheim, Norway.}

\author{Justin W. Wells}
\email[Corresponding author: ]{j.w.wells@fys.uio.no}
\affiliation{Department of Physics, Norwegian University of Science and Technology (NTNU), NO-7491 Trondheim, Norway.}
\affiliation{Centre for Materials Science and Nanotechnology, University of Oslo, Oslo 0318, Norway.}

\author{Federico Mazzola}
\affiliation{Istituto Officina dei Materiali (IOM)-CNR, Area Science Park-Basovizza, S.S. 14 Km 163.5, 34149 Trieste, Italy.}

\date{\today}

\begin{abstract}
Vicinal surfaces of bismuth are unique test-beds for investigating one-dimensional (1D) spin-polarised surface states that may one day be used in spintronic devices.  
In this work, two such states have been observed for the (112) surface when measured using angle-resolved photoemission spectroscopy (ARPES) and spin-resolved ARPES, and when calculated using a tight-binding (TB) model and with density functional theory (DFT). 
The surface states appear as elongated Dirac-cones which are 1D and almost dispersionless in the $\vb{k}_{\text{y}}$-direction, but disperse with energy in the orthogonal $\vb{k}_{\text{x}}$-direction 
to form two ``$\times$''-like features centered at the $\vb{k}_{\text{y}}$-line through $\Bar{\Gamma}$.  
Unlike many materials considered for spintronic applications, their 1D nature suggests that conductivity and spin-transport properties are highly dependent on direction.
The spin-polarisation of the surface states is mainly in-plane and parallel to the 1D state, but there are signs of a tilted out-of-plane spin-component for one of them. 
The Bi(112) surface states resemble those found for other vicinal surfaces of bismuth, strongly indicating that their existence and general properties are robust properties of vicinal surfaces of bismuth.
Furthermore, differences in the details of the states, particularly related to their spin-polarisation, suggest that spin-transport properties may be engineered simply by precise cutting and polishing of the crystal. 

\end{abstract}

\maketitle

\section{Introduction}\label{sec:intro}

To find materials where spin-filtering and spin-transport can be realised, either through electrical currents or magnon currents, are of major importance in spintronics \cite{Hirohata2020,Manipatruni2019,Bruene2012,Koenig2007, Roushan2009,Koo2009}. 
Spin states facilitating the flow of spin-polarised electrical currents are typically found in heavy element materials like bismuth, because of their strong spin-orbit coupling \cite{bawden2015hierarchical, Hofmann2006surfaces, noguchi2019weak,autes2016novel}.   

Two surfaces of bismuth, Bi(441) and Bi(114), have previously been found to support one-dimensional (1D), spin-polarised surface states \cite{Bianchi2015one,Wells2009nondegenerate}.
The one-dimensionality of these states suggests that conductivity and spin-transport can vary along different directions on the surface. This distinguishes them from many topological insulators for which conductivity and spin-transport properties tend to be closer to isotropic \cite{Hsieh2009, Xia2009, Chen2009, Pan2011}. 
Furthermore, the spin-polarisation of the surface states on Bi(441) and Bi(114) makes back-scattering of surface state electrons less likely since this implies a full reversal of their spin vectors \cite{Bruene2012, Koenig2007, Roushan2009}.
Another advantage of pure bismuth is how easy it is to cut, polish and clean compared to many topological insulators, which tend to be alloys \cite{Hsieh2009, Xia2009, Chen2009, Pan2011}. 
True vicinal surfaces of alloys are challenging to achieve consistently, as the surface stoichiometry and therefore the electronic properties will easily vary upon re-preparation.

One of the reasons for the intense interest in topological insulators is that the topology of the surface states protect electrons in these states from scattering, unless the scattering involves a mechanism that allows exchange of spin angular momentum \cite{Hsieh2009, Bruene2012,Koenig2007, Roushan2009, Xia2009, Chen2009, Pan2011}. Since topological insulators are ideally insulators in the bulk, electrical transport happens through their surface states. 
Bismuth is a semi-metal, so transport can happen through bulk states, where electrons are not generally protected from scattering, in addition to through surface states \cite{Wells2008, Hofmann2006surfaces}. 
However, if gaps could be opened in the energy spectrum of bismuth such that only surface states contribute to the Fermi-surface, or alternatively if vicinal surfaces could be achieved on similar topological insulators like $\mathrm{Bi}_{1-x}\mathrm{Sb}_{x}$ \cite{Guo2011, Roushan2009, Zhu2014}, they would be prime candidates for anisotropic electrical spin-transport. 
Vicinal surfaces of bismuth are therefore great test-beds to investigate the properties of such surface states, and understanding the underlying physical mechanisms should aid the exploration of multi-functional spin-based devices in the future. 

Here we show that another vicinal surface of bismuth, Bi(112), exhibits 1D, spin-polarised surface states resembling those found on Bi(441) and Bi(114) \cite{Bianchi2015one,Wells2009nondegenerate}.
Angle-resolved photoemission spectroscopy (ARPES) measurements reveal two spin-polarised ``$\times$''-like features in the energy spectrum, centered at $\Bar{\Gamma}_{1}$ in the centre of the 1st Brillouin zone (BZ, see Fig. \ref{fig:fig1}(d)), with each of their crossing points at approximately the same binding energies ($E_{\text{B}}$, also known as $E-E_{\text{F}}$) as the surface states of Bi(441) \cite{Bianchi2015one}. Similarly, Bi(114) also has a 1D spin-polarised surface state at one of these $E_{\text{B}}$ \cite{Wells2009nondegenerate}. Spin-polarised surface states therefore appear to be a robust property of vicinal bismuth surfaces.
Interestingly, all three surfaces show subtle differences in the details of the surface states, suggesting that properties like the direction of the spin-vector can be tuned simply by cutting the crystal along a given direction. This knowledge can possibly be exploited to create easily modifiable materials for use in new spintronic devices.  

\section{Results \& Discussion}

\begin{figure}
    \centering
    \includegraphics[width=0.45\textwidth]{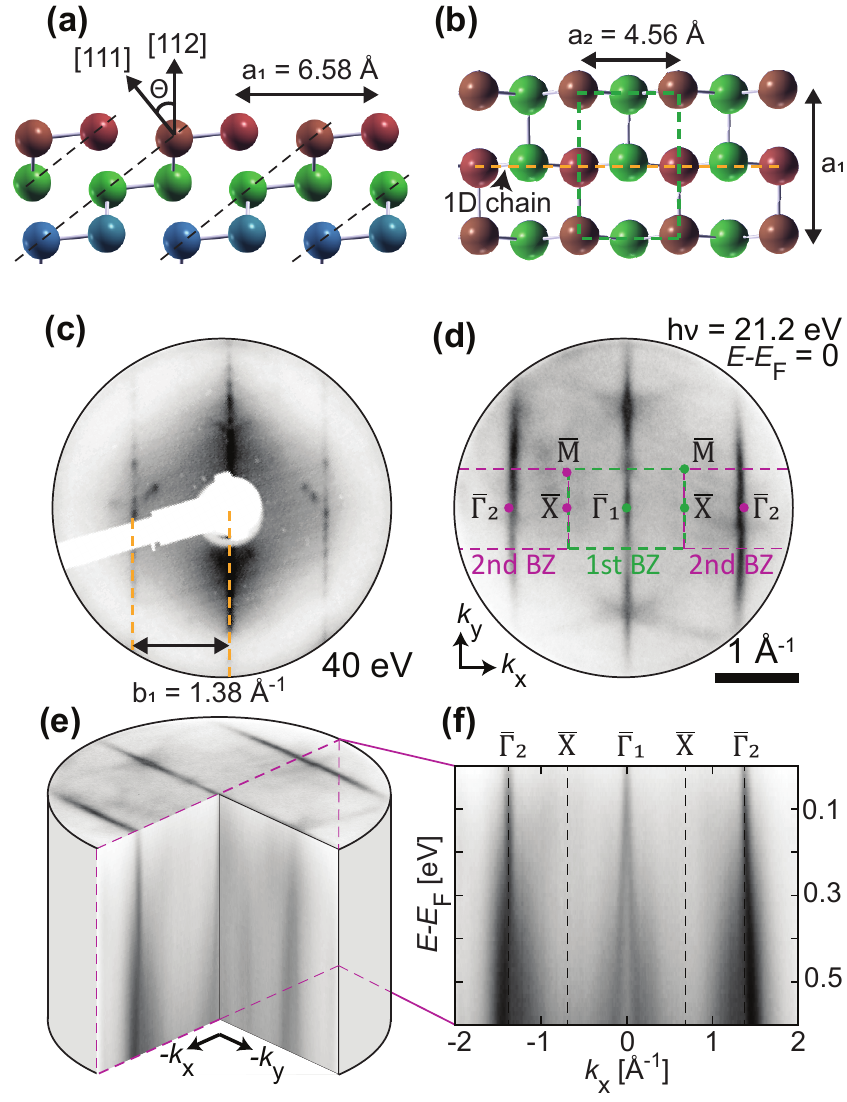}
    \caption{The atomic and electronic structure of Bi(112). \textbf{(a)} Sketch showing the atomic structure of the vicinal Bi(112) surface from the side. Red, green and blue spheres represent the uppermost, 2nd and 3rd layers of the crystal respectively. \textbf{(b)} Top view schematic of the Bi(112) surface with the unit cell indicated (green). \textbf{(c)} Low-energy electron diffraction (LEED) pattern of the Bi(112) surface, measured with electron energy 40~eV. \textbf{(d)} The two-dimensional Fermi-surface of Bi(112). The 1st and 2nd Brillouin zones (BZ) are indicated by green and purple rectangles. \textbf{(e)} Volumetric representation of the measured Bi(112) band structure.  \textbf{(f)} Measured band structure ($E$ vs $\vb{k}_{\text{x}}$) at $\vb{k}_{\text{y}}=0$, i.e. orthogonal to the one-dimensional surface states.}
    \label{fig:fig1}
\end{figure}

The vicinal surface of Bi(112) can be described by sheets of atoms with edges forming parallel lines on the surface (see Fig. \ref{fig:fig1}(a) and (b)). Due to dimerisation, the lines are only weakly coupled. This results in a macroscopic number of parallel quasi 1D systems. The surface structure shown in Figs. \ref{fig:fig1}(a) and (b) is confirmed by low-energy electron diffraction (LEED) in Fig. \ref{fig:fig1}(c). The distance between the streaks indicated in Fig. \ref{fig:fig1}(c) matches the periodicity of the truncated bulk crystal along the direction orthogonal to the 1D atomic chains. 
ARPES measurements further confirm the 1D character of the surface localised electrons. 
The Fermi-surface with 1D states along $\vb{k}_{\text{y}}$ is shown in  Fig. \ref{fig:fig1}(d). 
As seen in Figs. \ref{fig:fig1}(e) and (f), these lines consist of two bands in each BZ dispersing in energy to form ``$\times$''-like features, crossing each other at the $\Bar{\Gamma}$-points close to the Fermi-level. The bands resemble the shape of a Dirac cone, but with the Dirac point elongated to form a ``Dirac line'' parallel to $\vb{k}_{\text{y}}$ and an ``$\times$'' in the orthogonal direction along $\vb{k}_{\text{x}}$. 
Additional features in the measured spectrum are identified as bulk bands by comparison to density functional theory (DFT) calculations, see details in the Supplementary Material \cite{Suppl_Mat}. 

\begin{figure*}[t]
    \centering
    \includegraphics[width=0.95\textwidth]{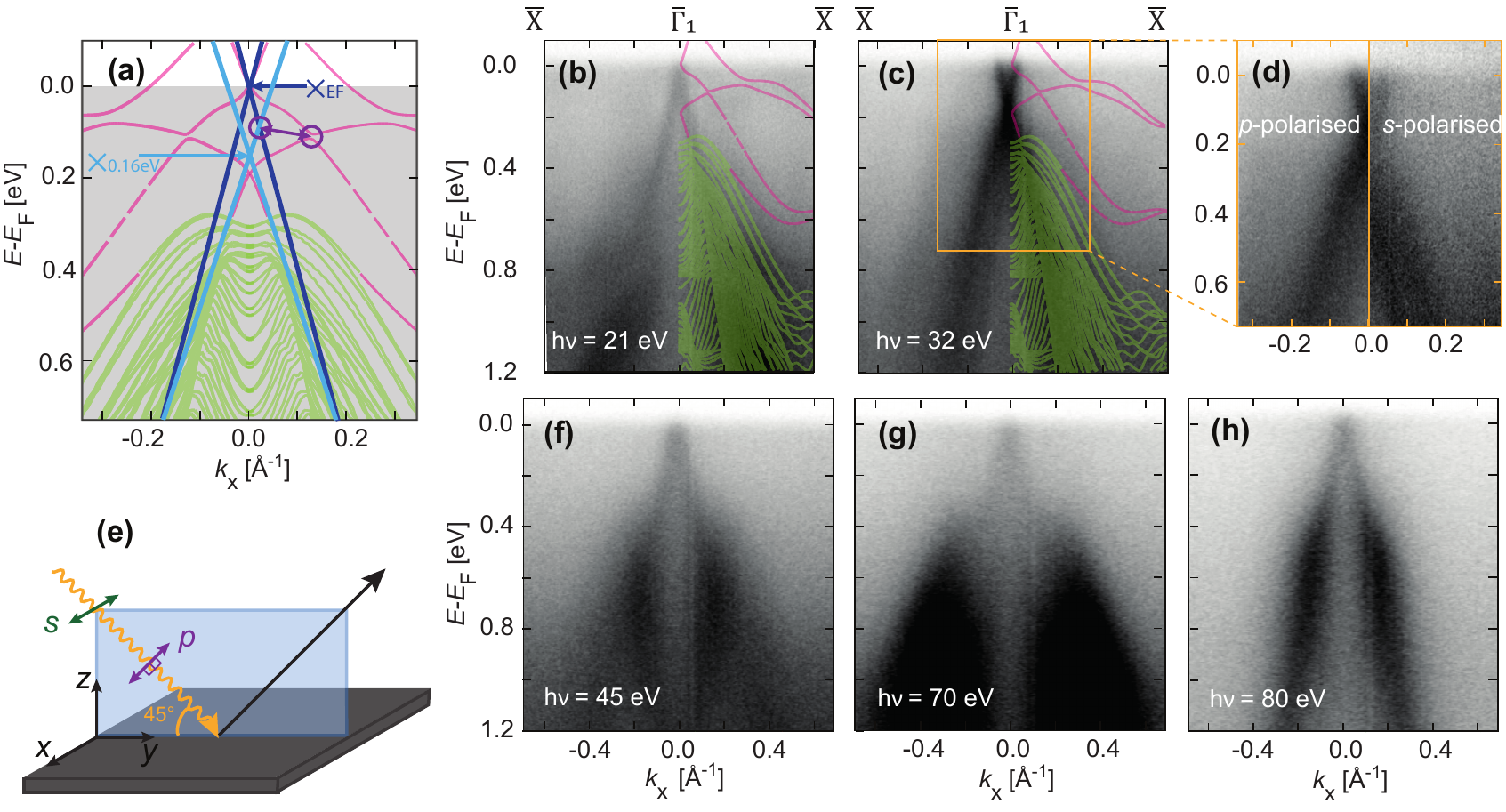}
    \caption{\textbf{(a)} Overview of the band structure near the $\Bar{\Gamma}_1$-point ($E$ vs $\mathbf{k_{\text{x}}}$) at $\mathbf{k_{\text{y}}}=0$. Green and pink bands are bulk and surface states, respectively, calculated using a tight-binding (TB) model. Dark blue and light blue bands represent the two ``$\times$''-like features at $\Bar{\Gamma}_1$ as seen in the measurements. Arrows from $\times_{E_\text{F}}$ and $\times_{0.16\,\text{eV}}$ indicate the crossing points of the $\times$'s. The purple circles and arrow show where $\times_{E_\text{F}}$ cross $\times_{0.16\,\text{eV}}$ away from $\Bar{\Gamma}_1$ in the measurements (dark blue cross light blue) and in the calculations (pink). \textbf{(b)}-\textbf{(d)}, \textbf{(f)}-\textbf{(h)} Band structure along $\Bar{\mathrm{X}}-\Bar{\Gamma}_1-\Bar{\mathrm{X}}$ measured at photoexcitation energies $h\nu=21$\,eV \textbf{(b)}, $h\nu=32$\,eV \textbf{(c)}-\textbf{(d)}, $45$\,eV \textbf{(f)}, $70$\,eV \textbf{(g)} and $80$\,eV \textbf{(h)}. TB calculations are overlaid on the measurements. In \textbf{(d)}, which is measured inside the orange region in \textbf{(c)}, the left hand side is measured with $p$-polarised light, and the right hand side with $s$-polarised light. The direction of the light polarisation with respect to the sample surface is shown in \textbf{(e)}.}
    \label{fig:Diffhv}
\end{figure*}

In order to differentiate surface states from bulk states, measurements were performed as a function of photoexcitation energy $h\nu$. Although symmetry breaking in the out-of-plane direction ($\hat{\vb{z}}$) caused by the surface means that the out-of-plane momentum ($\propto\vb{k}_{\text{z}}$) is not well conserved in the photoemission process, the  $\vb{k}_{\text{z}}$ dispersion of emitted bulk state electrons can still be probed by varying the excitation energy \cite{Damascelli2004}. On the other hand, surface states are localised on the surface and their energy is therefore independent of $\vb{k}_{\text{z}}$ (and $h\nu$) \cite{kevan1992angle}. 

An overview of the results is given in Fig. \ref{fig:Diffhv}(a), showing a sketch of the measured surface states (dark blue and light blue), compared to calculated tight-binding (TB) surface (pink) and bulk (green) states.
In Figs. \ref{fig:Diffhv}(b)-(d) and (f)-(h), the bands at $E_{\text{B}}$ larger than $\approx 0.3$\,eV are seen to disperse with photoexcitation energy and are therefore identified as bulk bands. In addition, there are two ``$\times$''-like features formed by bands crossing each other at $\Bar{\Gamma}_{1}$, i.e. $\times_{E_\text{F}}$ and  $\times_{0.16\,\text{eV}}$, shown as dark blue and light blue bands respectively in Fig. \ref{fig:Diffhv}(a). These bands do not disperse with photoexcitation energy, hence they are identified as surface states. However, a noticeable variation in intensity is observed, most likely related to matrix element effects \cite{Damascelli2004, Day2019}. As indicated by arrows in Fig. \ref{fig:Diffhv}(a), $\times_{E_\text{F}}$ (dark blue) has its crossing point near the Fermi-level at $E_\text{B}=0.07 \pm 0.10$\,eV. 
Whereas, $\times_{0.16\,\text{eV}}$ (light blue in Fig. \ref{fig:Diffhv}(a)) has the crossing point at a higher $E_\text{B}$ of $E_\text{B}=0.16 \pm 0.05$\,eV, and can be seen from the measurements in Fig. \ref{fig:Diffhv}(c). See the Supplementary Material for details on how the $E_{\text{B}}$ of the crossing points were determined \cite{Suppl_Mat}. 

The electronic structure of Bi(112) was calculated using TB and DFT. Results from the DFT-calculations are included in the Supplementary Material \cite{Suppl_Mat}. The TB calculations are underlying the $\times$'s in Fig. \ref{fig:Diffhv}(a) and overlaid on the measured band structures in Figs. \ref{fig:Diffhv}(b) and (c). The bulk states (green) have been calculated for a range of $\vb{k}_\text{z}$, hence the measured bulk states are expected to disperse within the area they cover. There is good agreement between the TB calculated (green) and measured bulk states.  

The TB calculated surface states (pink) show band structure similar to the measurement in terms of two $\times$-like features with crossing points at $E_\text{B}=-0.032$\,eV and $E_\text{B}=0.166$\,eV for $\times_{E_\text{F}}$ and $\times_{0.16\,\text{eV}}$, respectively. 
Although the $E_{\text{B}}$ of the crossing points in the measurements agree within uncertainties with the values expected from calculations, the gradients of the bands differ.
Below the $\times_{E_\text{F}}$ crossing point and above the $\times_{0.16\,\text{eV}}$ crossing point, the measured bands are much steeper than the calculated bands. Because of this, the point where the bands extending from $\times_{E_\text{F}}$ and $\times_{0.16\,\text{eV}}$ are measured to cross (purple circles and arrow in Fig. \ref{fig:Diffhv}(a)) is at a different value of $\vb{k}_{\text{x}}$ and $E_\text{B}$ compared to the calculated bands. Preliminary calculations indicated that surface state gradients and energies are very sensitive to the precise geometry at the surface \cite{2022JBMasters}. 
Relaxation of the surface, or missing rows of atoms as observed in Ref. \onlinecite{Wells2009nondegenerate}, may therefore explain the difference in band-gradients, potentially leading to a mechanism to control the surfance Fermi velocities. 
Such missing rows may also explain why the surface states appear more 1D than expected from the calculations: if the distance between the sheet edges is greater, a weaker interaction between their constituent electronic states is expected \cite{Crain2004}. 

\begin{table}[ht]
    \centering
    \begin{threeparttable}[b]
    \caption{Comparison of the binding energy ($E_{\text{B}}$) of the crossing points and spin-vector of the $\times$-like features in the measured band structure of three surfaces of bismuth. TB Bi(112) gives the equivalent results from tight-binding calculations. The spin-vector angle gives the angle between the spin-vector and the surface plane, i.e. a measure of how large the out-of-plane spin-component is.}
    \label{tab:CompCross}
    \begin{tabular}{lcccc}
        \hline
        Surface & \multicolumn{2}{c}{$E_\text{B}$ Crossing Point [eV]} & \multicolumn{2}{c}{Spin-Vector Angle [$^{\circ}$]}  \\
         & $\times_{E_\text{F}}$ & $\times_{0.16\,\text{eV}}$ &  $\times_{E_\text{F}}$ & $\times_{0.16\,\text{eV}}$ \\ \hline
        Bi(112) & $0.07 \pm 0.10$ & $0.16 \pm 0.05$ & $0$ & $27 \pm 7$ \\ 
        Bi(441) \cite{Bianchi2015one} & $\approx 0$ & $ 0.15-0.20$ & $0$ & $45$ \\
        Bi(114) \cite{Wells2009nondegenerate} & $0-0.15$\tnote{a} & Not seen & $30$ & Not seen  \\ 
        TB Bi(112) & $-0.032 $ & $0.166 $  & $\approx 0$ & $\approx 0$  \\ 
       \hline 
    \end{tabular}
    \begin{tablenotes}
    \item [a] For Bi(114), the surface state looks like a single line feature on top of a $\Lambda$-like bulk state, hence it is not clear whether the surface state is $\times$-like with a crossing point. The maximum $E_{\text{B}}$ in the given range is estimated as the highest $E_{\text{B}}$ of the surface state. See details in Ref.~\onlinecite{Wells2009nondegenerate}.
    \end{tablenotes} 
    \end{threeparttable}
\end{table}

The $\times$-like features are similar to the surface states observed for other vicinal surfaces of bismuth, as summarised in Table \ref{tab:CompCross}. Two $\times$-like surface states with their crossing point near the Fermi-level and at $E_\text{B}=0.16 -0.20$\,eV are observed for both the (112) and (441) surface \cite{Bianchi2015one}. The band structure of Bi(114) has a $\Lambda$-shaped bulk feature with its maximum point near $E_\text{B}\approx 0.15$\,eV, but with an additional 1D surface state at $E_{\text{B}}$ between $0-0.15$\,eV \cite{Wells2009nondegenerate}. For all three surfaces, the one-dimensionality of the surface states is expected to influence the conduction properties. A distinct increase in the conductivity along the atomic rows of the surface when compared to the orthogonal direction is expected as there are far more states to scatter into along the surface chains (see Fig. \ref{fig:fig1}(b)).

In order to investigate the orbital origins of the observed states, the polarisation of the light was varied between $s$- and $p$- polarisation (in-plane and partly out-of-plane respectively) as visualised in figure \ref{fig:Diffhv}(e).
An example is given in Fig. \ref{fig:Diffhv}(d), where an additional band can be observed inside the $\times_{0.16\,\text{eV}}$-feature when measured with $s$-polarised light (right hand side panel). Since this is in the region where bulk bands are found from the calculations, it is believed to be a bulk band. See the Supplementary Material for more details on the orbital nature of the states \cite{Suppl_Mat}.  

\begin{figure}
    \centering
    \includegraphics[width=0.45\textwidth]{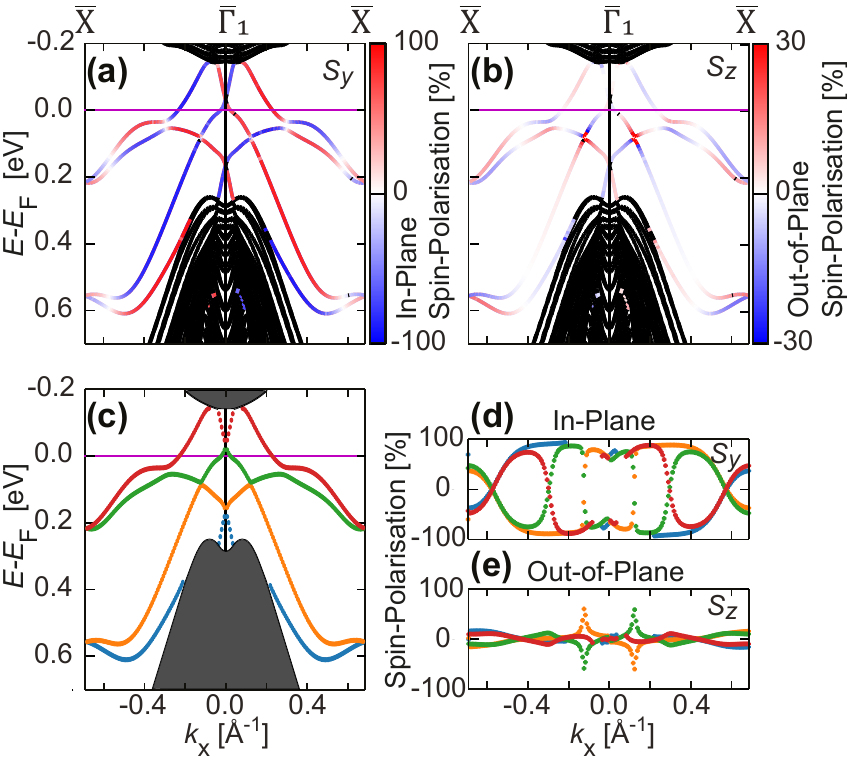}
    \caption{Spin-polarisation from tight-binding calculations. \textbf{(a)}, \textbf{(b)}, \textbf{(c)} Band structure between $\Bar{\text{X}}-\Bar{\Gamma}_{1}-\Bar{\text{X}}$. In \textbf{(a)} and \textbf{(b)} the colour-scale shows the spin-polarisation $\vb{S}_{\text{y}}$, the spin-vector component in-plane and along the 1D lines, \textbf{(a)}, and $\vb{S}_{\text{z}}$, the out-of-plane component of the spin-vector, \textbf{(b)}. In \textbf{(c)}, the colours label the surface bands. \textbf{(d)}, \textbf{(e)} Spin-polarisation of the spin-vector component $\vb{S}_{\text{y}}$ \textbf{(d)} and $\vb{S}_{\text{z}}$ \textbf{(e)} respectively. The colour shows which band in \textbf{(c)} the spin-polarisation value belongs to.}
    \label{fig:spinCalc}
\end{figure}

Due to the strong spin-orbit coupling in bismuth, the surface states are expected to be spin-polarised \cite{Hofmann2006surfaces,bawden2015hierarchical}. 
This is demonstrated in Fig. \ref{fig:spinCalc}, 
showing the TB calculated spin-polarisation dependence of the surface states. 
Positive (negative) spin-polarisation shown in red (blue) gives the magnitude of the spin-vector component in-plane and pointing along $+\vb{k}_{\text{y}}$ ($-\vb{k}_{\text{y}}$), $\vb{S}_{\text{y}}$, in Fig. \ref{fig:spinCalc}(a) and out-of-plane pointing along $+\vb{k}_{\text{z}}$ ($-\vb{k}_{\text{z}}$), $\vb{S}_{\text{z}}$, in Fig. \ref{fig:spinCalc}(b). 
For all the surface bands, $\vb{S}_{\text{y}}$ is close to $\pm100$\,\% (Figs. \ref{fig:spinCalc}(a) and (d)). The exception is the points where the surface states corresponding to the two crosses $\times_{0.16\,\text{eV}}$ and $\times_{E_\text{F}}$ cross (purple circles and arrow in Fig. \ref{fig:Diffhv}(a)). There, the surface states also attain a significant out-of plane spin-component $\vb{S}_{\text{z}}$ (Figs. \ref{fig:spinCalc}(b) and (e)).

\begin{figure}
    \centering
    \includegraphics[width=0.45\textwidth]{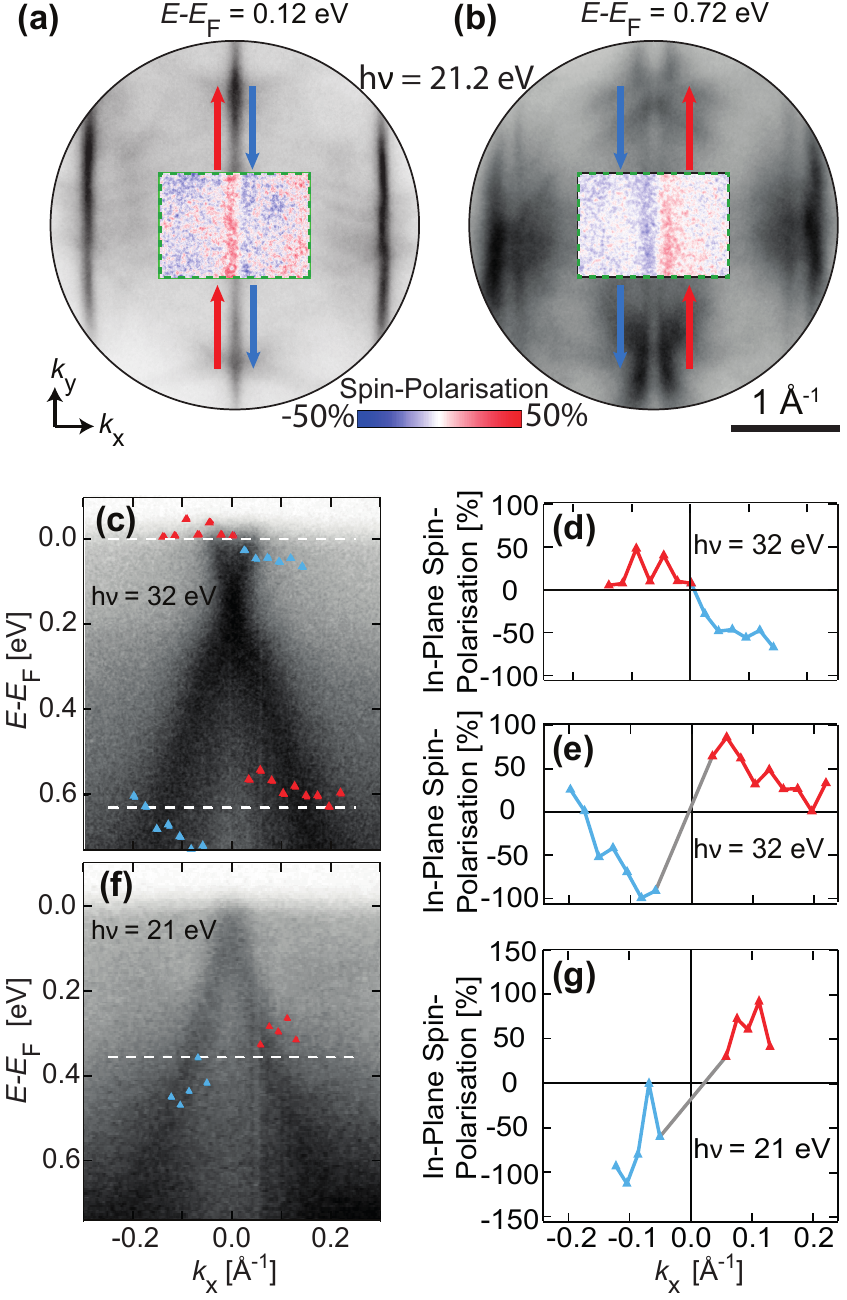}
    \caption{Measured spin texture. \textbf{(a)}, \textbf{(b)} Constant energy surfaces at binding energies $E_\text{B}=0.12$\,eV \textbf{(a)} and $E_\text{B}=0.72$\,eV \textbf{(b)}. The first Brillouin zone is substituted by the in-plane spin-polarisation along the $\vb{k}_{\text{y}}$-direction visualised by red and blue arrows. \textbf{(c)}, \textbf{(f)} $E$ vs $\vb{k}_{\text{x}}$ plot measured with photoexcitation energies of $h\nu=32$\,eV and $h\nu=21$\,eV. The in-plane spin-polarisation along $\vb{k}_{\text{y}}$ was measured along the dashed white lines and has been overlaid. \textbf{(d)}, \textbf{(e)}, \textbf{(g)} In-plane spin-polarisation along $\vb{k}_{\text{y}}$ at $E_\text{B}=0$\,eV \textbf{(d)} and $E_\text{B}=0.63$\,eV \textbf{(e)} in \textbf{(c)}, and $E_\text{B}=0.36$\,eV \textbf{(g)} in \textbf{(f)}. 
    }
    \label{fig:SpinMeas}
\end{figure}

The measured in-plane spin-polarisation shown in Fig. \ref{fig:SpinMeas} is consistent with the calculations. 
In Figs. \ref{fig:SpinMeas}(a) and (b), constant energy surfaces are shown at $E_\text{B}=0.12$\,eV and $E_\text{B}=0.72$\,eV. 
At both energies, the region spanned by the 1st BZ (green, dashed rectangle) have been filled with the measured projection of the spin-vector in-plane and along $\vb{k}_{\text{y}}$  ($\vb{S}_{\text{y}}$).
At $E_\text{B}=0.12$\,eV, the positive spin-polarisation (red) occurs for negative values of $\vb{k}_{\text{x}}$, while at $E_\text{B}=0.72$\,eV, the positive spin-polarisation is at positive values of $\vb{k}_{\text{x}}$. This indicates that the spin-polarisation reverses at a band crossing between $E_\text{B}=0.12-0.72$\,eV. The spin-reversal is confirmed in Figs. \ref{fig:SpinMeas}(c) and (f) where the in-plane spin-polarisation $\vb{S}_{\text{y}}$ along the 1D line is overlaid on the $E$ vs $\vb{k}_{\text{x}}$ maps of the surface states. $\vb{S}_{\text{y}}$ was found to be $\approx30$\,\% when measured above and below the crossing point for $\times_{0.16\,\text{eV}}$, and $\approx70$\,\% for $\times_{E_\text{F}}$, see Figs. \ref{fig:SpinMeas}(d), (e) and (g), respectively. 

It should be noted that the spin-texture is similar for both $\times_{E_\text{F}}$ and $\times_{0.16\,\text{eV}}$ although their magnitude differ, as seen when comparing Figs. \ref{fig:SpinMeas}(e) and (g). 
Because of this, we presume that backscattering is strongly reduced for electronic states in the direction corresponding to the atomic chains in Fig. \ref{fig:fig1}(b).  
In addition, the spin-polarisation means that instabilities like charge density waves are not expected on the Bi(112) surface \cite{Kim2005, Leuenberger2013}. 

Signs of an out-of-plane spin-vector component can also be seen when measuring the spin-polarisation of $\times_{0.16\,\text{eV}}$  (see details in the Supplementary Material \cite{Suppl_Mat}). However, this spin-component is small compared to $\vb{S}_{\text{y}}$, and approaches the detection limit in our measurements. 
Even though $\vb{S}_{\text{z}}$ is small in the measurements and is only significant for short $\vb{k}_{\text{x}}$-ranges in the calculations, it can explain why the measured $\vb{S}_{\text{y}}$ in Fig. \ref{fig:SpinMeas}(d) is lower than expected from the calculations. These spin measurements were performed near the Fermi-level, which is close to where $\times_{0.16\,\text{eV}}$ and $\times_{E_\text{F}}$ cross (purple circles and arrow in Fig. \ref{fig:Diffhv}(a)). At this point, $\vb{S}_{\text{z}}$ increases drastically in the calculations, and consequently $\vb{S}_{\text{y}}$ decreases. 

When comparing the spin-polarisation found for Bi(112) with Bi(441) and Bi(114) (see Table \ref{tab:CompCross}), they are found to be similar, but with important differences \cite{Bianchi2015one,Wells2009nondegenerate}. All three surfaces have 1D surface states where the in-plane spin-component along the 1D states ($\vb{S}_{\text{y}}$) is the strongest component. For $\times_{E_\text{F}}$ in Bi(112) and Bi(441) this is the only component. 
$\times_{E_\text{F}}$ in Bi(114) and $\times_{0.16\,\text{eV}}$ in Bi(441) was previously found to have an out-of-plane component in addition to the in-plane component, such that the spin-vector makes an angle $\approx 30^\circ$ and $\approx45^\circ$ with the (114) and (441) surface plane respectively \cite{Bianchi2015one,Wells2009nondegenerate}. 
Bianchi \textit{et al.} (2015) \cite{Bianchi2015one} suggested that the out-of-plane component of the spin may be a consequence of exposing edges of the (111) plane in bismuth (see Fig. \ref{fig:fig1}(a)), and that this spin-vector is in fact in-plane in the (111) plane. This would mean that $\times_{E_\text{F}}$ in Bi(114) and $\times_{0.16\,\text{eV}}$ in Bi(441) correspond to surface states in Bi(111), while $\times_{E_\text{F}}$ in Bi(441) corresponds to surface states on the (441) surface.
This hypothesis seems to match well for Bi(441) and Bi(114) \cite{Bianchi2015one,Wells2009nondegenerate} when comparing the angle of the spin-vector with the angle between the (111) and (114), or (441) surface planes. 

To see whether $\times_{E_\text{F}}$ and $\times_{0.16\,\text{eV}}$ in Bi(112) also can be related to states on different surfaces, the full spin-vector was estimated. 
As mentioned earlier, the spin-vector of $\times_{E_\text{F}}$ seems to be fully in-plane, hence indicating that it belongs to a surface state on Bi(112). $\times_{0.16\,\text{eV}}$ on the other hand, seems to have a small out-of-plane component, causing the spin-vector to be tilted out-of-plane to make an angle of $(27 \pm 7)^\circ$ with the surface plane. This angle is smaller than the angle of $37^\circ$ between the (111) and (112) plane, but almost within the uncertainty of the estimate. 
Altogether, the three surfaces have similar spin-vectors, but the angle between the surface plane and spin-vector seems to vary. These subtle but important differences mean that the spin-vector potentially can be tuned by choosing a particular vicinal surface.

\section{Conclusion}
One-dimensional, spin-polarised surface states have been observed on the (112) surface of bismuth. The surface states are seen as ``$\times$''-like features in the band structure with crossing points at $E_\text{B}=0.07 \pm 0.10$\,eV and $E_\text{B}=0.16 \pm 0.05$\,eV for $\times_{E_\text{F}}$ and $\times_{0.16\,\text{eV}}$ respectively. These values agree within uncertainties with calculations, even if the gradients of the measured bands differ from the calculations. 
The spin-vector of the surface states is mainly in-plane and along the 1D lines, but $\times_{0.16\,\text{eV}}$ shows indications of having an additional out-of-plane spin-component, approximately matching with the angle of the (111) plane. The surface states observed in Bi(112) resemble those found for other vicinal surfaces of bismuth, indicating that the existence of such states is a robust property of these surfaces. Furthermore, variations in how the spin-vector is tilted out-of-plane for different surfaces, can make it possible to tune the spin-transport on bismuth-surfaces by adjusting the surface crystallographic direction.

\section{Acknowledgements}

This work was partly supported by the Research Council of Norway, project numbers 324\,183, 315\,330, and 262\,633. We acknowledge Elettra Sincrotrone Trieste for providing access to its synchrotron radiation facilities and for financial support under the IUS internal project (proposal 20215677).
This work has been partly performed in the framework of the nanoscience foundry and fine analysis (NFFA-MUR Italy Progetti Internazionali) facility.

\section*{Appendix: Methods}

\subsection{Sample Preparation}
A clean Bi(112) surface was prepared by repeated cycles of Ar$^+$ ion sputtering at $200$-\SI{400}{\eV}, followed by annealing to $T\approx\SI{50}{\celsius}$ for short duration. 
The cleanliness of the surface was verified by sharp and oxide free Bi core levels using X-ray photoelectron spectroscopy (XPS), and the crystallinty of the surface by LEED.

\subsection{Band Structure and Spin Measurements}

\subsubsection{Momentum Microscopy Measurements}
Band structure measurements were performed at $T\approx\SI{115}{\kelvin}$ using a NanoESCA III aberration corrected EF-PEEM equipped with a focused helium discharge lamp primarily generating He I photons at $h\nu= 21.2$\,eV, using pass energy $E_{\text{P}}=\SI{25}{\eV}$ and a \SI{1.0}{\mm} entrance slit to the energy filter. With the given settings, the instrument had nominal energy and momentum resolutions of approximately \SI{100}{\meV} and $0.02~\text{Å}^{-1}$, respectively. 

Two-dimensional spin-polarised measurements at constant energies were performed using an Ir spin filter coated with a monolayer of Au  \cite{Tusche2015, Giebels2013}. The spin filter measured the projection of the spin along $\pm \vb{k}_{\text{y}}$, in-plane and along the 1D-line. The spin-polarisation $P$ was calculated from the measurements using 

\begin{equation}
    P= \frac{I_{\uparrow}-I_{\downarrow}}{S(I_{\uparrow}+I_{\downarrow})}
\end{equation}

\noindent where $I_{\uparrow}$ and $I_{\downarrow}$ are the intensities of the energy surface when filtering spin along $+\vb{k}_{\text{y}}$ and $-\vb{k}_{\text{y}}$, respectively \cite{Meier2009measuring}. A Sherman function $S=0.6$ was assumed, based on preliminary calibration measurements \cite{Meier2009measuring, Sherman1956}.

\subsubsection{High-Resolution Measurements}
Higher energy resolution band structure measurements were performed at the APE-LE endstation at Elettra Synchrotron, Italy, using a VG SCIENTA DA30 analyzer whilst cooling the Bi(112) crystal to $T=\SI{77}{\kelvin}$. 

Spin measurements were performed using two three-dimensional vectorial spin-polarimeters operated in the very low energy electron diffraction (VLEED) regime \cite{Bigi2017very}. From the spin signals detected by the two VLEED spin detectors it is possible to reconstruct the full three-dimensional spin-vector carried by the emitted photoelectrons. A Sherman function of $S=0.3$ was found from calibration and was used in the analysis.

\subsection{Calculation Details}
\subsubsection{Tight-Binding Calculations}
The bulk band structure of bismuth was calculated within a TB model in Ref.~\onlinecite{Liu1995}. In this paper, we use the same model and parameters to calculate the band structure of a semi-infinite system with a finite number of layers in the $\hat{\vb{z}}$-direction and periodic boundary conditions in the $\vb{\hat{x}\hat{y}}$-plane, where the sample has been oriented so that the $\vb{\hat{x}\hat{y}}$-plane corresponds to the (112)-surface of bismuth. By performing the Fourier transform for the in-plane coordinates, the energy band structure can be calculated as eigenvalues of an effective 1D problem at every in-plane momentum $\vb{k}$. From the eigenvector corresponding to each eigenvalue at a given in-plane momentum, one may extract various properties of the state, such as orbital content, spin, and the spatial distribution in the direction perpendicular to the surface plane. The latter can be used to discriminate between between bulk and surface states, and is the basis for the choice of color in Fig. \ref{fig:Diffhv}. Within the geometry in the TB calculation, there are two surfaces, which both host surface states. In Fig. \ref{fig:spinCalc}, however, we only show the surface states corresponding to one of the two surfaces, while the opposing surface contains surface states with opposite spin-polarisation. 

\vspace{5mm}
\subsubsection{DFT Calculations}
First principles DFT calculations were carried out to better understand the electronic structure and to further confirm the validity of the TB model. All calculations were carried out with the QuantumESPRESSO DFT package using fully relativistic pseudopotentials and projector augmented wave exchange-correlation functionals (KJPAW). Bulk calculations were performed with a \SI{40}{\rydberg} plane wave cutoff and convergence threshold \SI{1e-8}{\rydberg}. The $\vb{k}$-points were sampled using a Monkhorst-Pack grid of $12\times12\times12$ and the lattice constant was found by a relaxation process. Surface states were calculated using a slab geometry with 24 atomic layers and a separation of \SI{15}{\angstrom} between slabs. Sampling of $\vb{k}$-points was done using a Monkhorst-Pack grid of $10\times10\times1$. The cut-off energy was \SI{40}{\rydberg} and the convergence threshold \SI{1e-6}{\rydberg}.

\bibliography{Bismuth}% Produces the bibliography via BibTeX.

\end{document}

\preprint{APS/123-QED}

\title{Supplementary Material for One-Dimensional Spin-Polarised Surface States - A Comparison of Bi(112) with Other Vicinal Bismuth Surfaces}

\author{Anna Cecilie Åsland}
\author{Johannes Bakkelund}
\affiliation{Department of Physics, Norwegian University of Science and Technology (NTNU), NO-7491 Trondheim, Norway.}

\author{Even Thingstad}
\affiliation{Department of Physics, Norwegian University of Science and Technology (NTNU), NO-7491 Trondheim, Norway.}
\affiliation{Department of Physics, University of Basel, Klingelbergstrasse 82, CH-4056 Basel, Switzerland.}

\author{Håkon I. Røst}
\affiliation{Department of Physics, Norwegian University of Science and Technology (NTNU), NO-7491 Trondheim, Norway.}

\author{Simon P. Cooil}
\affiliation{Centre for Materials Science and Nanotechnology, University of Oslo, Oslo 0318, Norway.}

\author{Jinbang Hu}
\affiliation{Department of Physics, Norwegian University of Science and Technology (NTNU), NO-7491 Trondheim, Norway.}

\author{Ivana Vobornik}
\author{Jun Fujii}
\affiliation{Istituto Officina dei Materiali (IOM)-CNR, Area Science Park-Basovizza, S.S. 14 Km 163.5, 34149 Trieste, Italy.}

\author{Asle Sudbø}
\affiliation{Department of Physics, Norwegian University of Science and Technology (NTNU), NO-7491 Trondheim, Norway.}

\author{Justin W. Wells}
\email[Corresponding author: ]{j.w.wells@fys.uio.no}
\affiliation{Department of Physics, Norwegian University of Science and Technology (NTNU), NO-7491 Trondheim, Norway.}
\affiliation{Centre for Materials Science and Nanotechnology, University of Oslo, Oslo 0318, Norway.}

\author{Federico Mazzola}
\affiliation{Istituto Officina dei Materiali (IOM)-CNR, Area Science Park-Basovizza, S.S. 14 Km 163.5, 34149 Trieste, Italy.}

\date{\today}% It is always \today, today,

\maketitle

\section{DFT-Calculations}
\label{sec:DFTSupp}
 
\begin{figure}[t]
    \centering
    \includegraphics[width=0.4\textwidth]{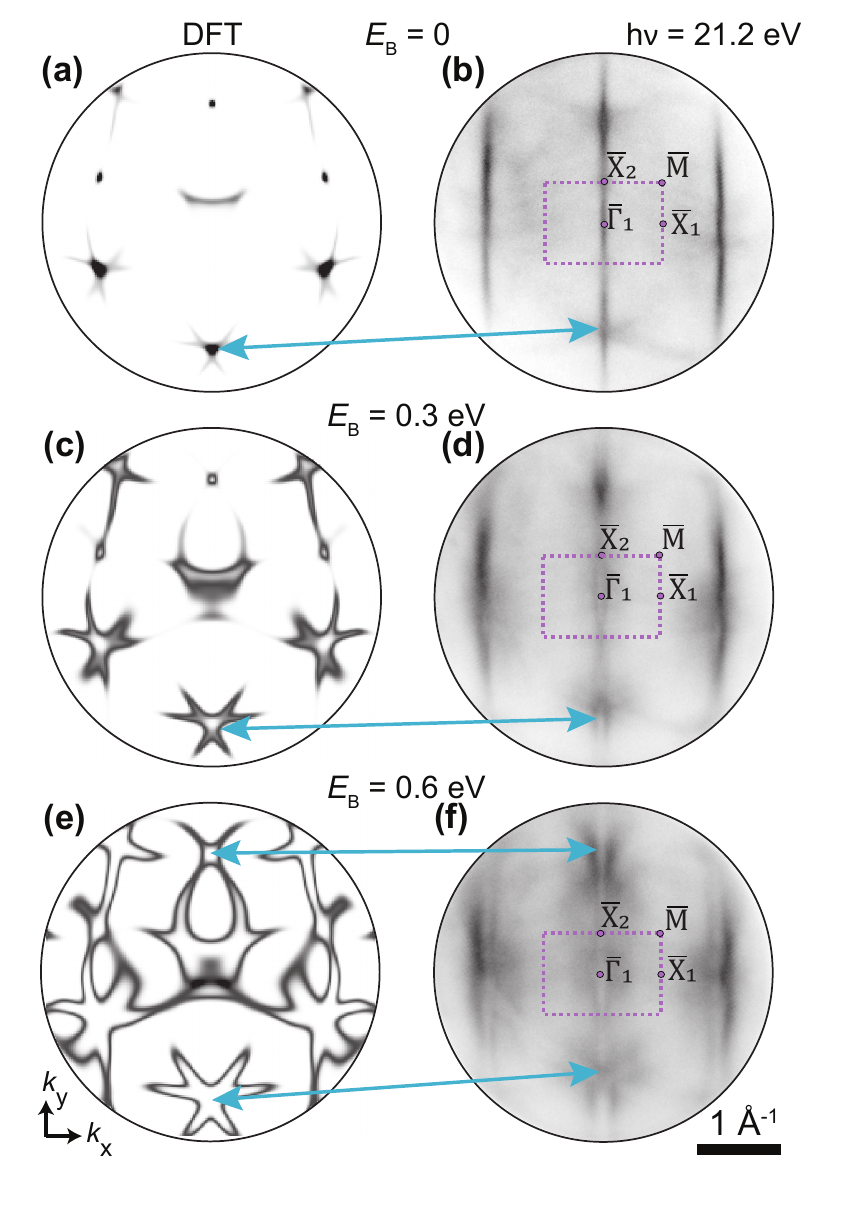}
    \caption{DFT-calculated constant energy surfaces (\textbf{(a)}, \textbf{(c)} and \textbf{(e)}) compared to constant energy surfaces measured at photoexcitation energy $h\nu=21$\,eV (\textbf{(b)}, \textbf{(d)} and \textbf{(f)}). Energy surfaces are shown at the Fermi-level ($E_{\text{B}}=0$\,eV), and $E_{\text{B}}=0.3$\,eV and $E_{\text{B}}=0.6$\,eV below the Fermi-level in \textbf{(a)}-\textbf{(b)}, \textbf{(c)}-\textbf{(d)} and \textbf{(e)}-\textbf{(f)}, respectively. The rectangles indicate the first Brillouin zone. Blue arrows indicate bulk states.}
    \label{fig:DFT_ES}
\end{figure}

Density functional theory (DFT) calculations were performed to determine the projected bulk band structure of Bi(112), together with surface states and their spin-polarisation (see Methods in the main text). Constant energy surfaces (ESs) were extracted from the calculations and compared to measured ESs in Fig. \ref{fig:DFT_ES}. The one-dimensional (1D) lines along $\vb{k}_{\text{y}}$ is not present in the DFT-calculations, hence they are thought to be surface states. Other features, some of them indicated by arrows, are bulk states. These can be seen in both the calculated and measured ESs, showing good agreement between the measurements and the calculations.  

\begin{figure}[t]
    \centering
    \includegraphics[width=0.45\textwidth]{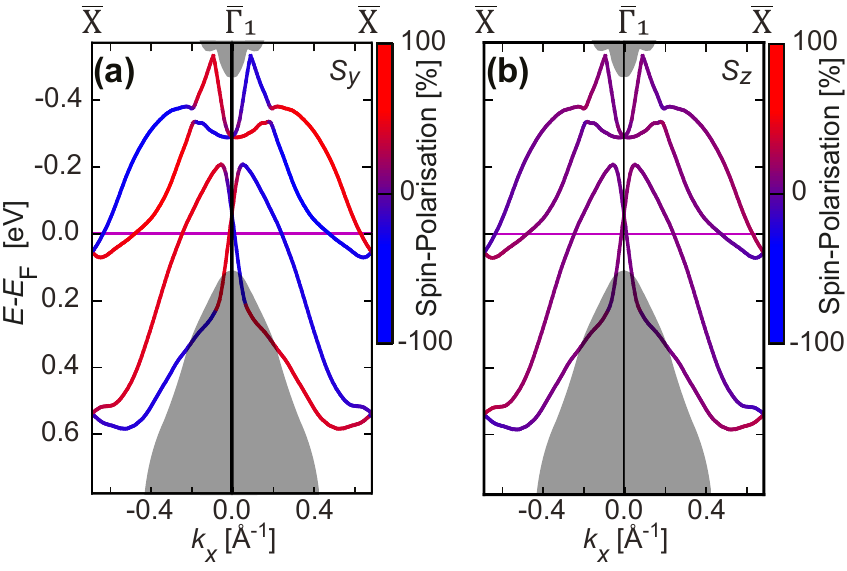}
    \caption{Band structure of calculated surface states using DFT. The colour-scale corresponds to the spin-polarisation $\vb{S}_{\text{y}}$, the spin-vector component in-plane and along the 1D lines, \textbf{(a)}, and $\vb{S}_{\text{z}}$, the out-of-plane component of the spin-vector, \textbf{(b)}.}
    \label{fig:DFT_bands}
\end{figure}

The DFT calculated surface states in Fig. \ref{fig:DFT_bands} show the same features and spin-polarisation as that found from tight-binding (TB) calculations and measurements (see Fig. 2 and 3 in the main text). The main difference is found in the binding energy ($E_{\text{B}}$) of the states. All the bands in the DFT calculations seem to be stretched compared to bands from TB calculations, causing the Fermi-level to be shifted towards lower $E_{\text{B}}$ by $E_{\text{B}}\approx 0.2$\,eV. The reason for this is unclear, but is maybe related to the thickness of the slab used in the calculations.  

\section{Crossing Points of the \texorpdfstring{$\times$}-like Surface States}
As seen in Fig. 2(a) in the main text, the surface states look like two $\times$-like features in the measurements. 
The crossing points of $\times_{E_\text{F}}$ and $\times_{0.16\,\text{eV}}$ are indicated by arrows in Fig. 2(a), and were found by fitting lines to the band-dispersions of the measured bands. The band-dispersions were determined by fitting a Voigt-function to the momentum distribution curves (MDCs) at each binding energy and plotting the $\vb{k}_{\text{x}}$-value of the peak as a function of binding energy. There are two bands in each $\times$, and the crossing point is set to be at the binding energy where the lines fitted to these bands cross each other at $\vb{k}_{\text{x}}=0$. 
For $\times_{E_\text{F}}$ the crossing point is at $E_\text{B}=0.07 \pm 0.10$\,eV and for $\times_{0.16\,\text{eV}}$ it is at $E_\text{B}=0.16 \pm 0.05$\,eV. 
Note that the uncertainty is larger than the energy resolution. The main origin of the uncertainty is some overlap between bulk and surface states below $E_\text{B}\approx0.2$\,eV, and assuming the bands in the $\times$'s to be linear, when in reality the gradient can vary, as seen from calculations.
The uncertainty is larger for $\times_{E_\text{F}}$ compared to $\times_{0.16\,\text{eV}}$, because the intensity of $\times_{E_\text{F}}$ is weaker and therefore more challenging to peak fit. 

\section{Orbital Origin of Electronic States}
The orbital origin of the electronic states may be probed by varying the light-polarisation of the incoming light in the ARPES measurements, since light with different polarisations couples to electrons in different orbitals \cite{Himpsel1980}. In the measurements presented in this work, two light-polarisations have been used: $s$-polarised and $p$-polarised light, see figure 2(e) in the main text. The analyser slit is aligned along the $\vb{\hat{x}}$-direction defined in figure 2(e).
Figs. \ref{fig:SpinSuppl}(a) and (b) are measured in the same way, but with $p$- and $s$-polarised light, respectively. The same $\times_{0.16\,\text{eV}}$-like feature can be observed in both cases, in agreement with the calculations.  
The additional bands seen inside $\times_{0.16\,\text{eV}}$ in Fig. \ref{fig:SpinSuppl}(b) are thought to be bulk bands with electrons from $p_x$-orbitals (because they are only seen with $s$-polarised light), as they are in the region where bulk bands are found.  

\section{Spin-Polarisation}
The spin-polarisation of the surface states from TB calculations is shown in Fig. 3 in the main text, and from DFT calculations in Fig. \ref{fig:DFT_bands}. The in-plane spin-component perpendicular to the 1D lines ($\vb{S}_{\text{x}}$) is not included, because $\vb{S}_{\text{x}}=0$ due to symmetry arguments. This was confirmed by measurements.  

\begin{figure*}[hbtp]
    \centering
    \includegraphics[width=0.675\textwidth]{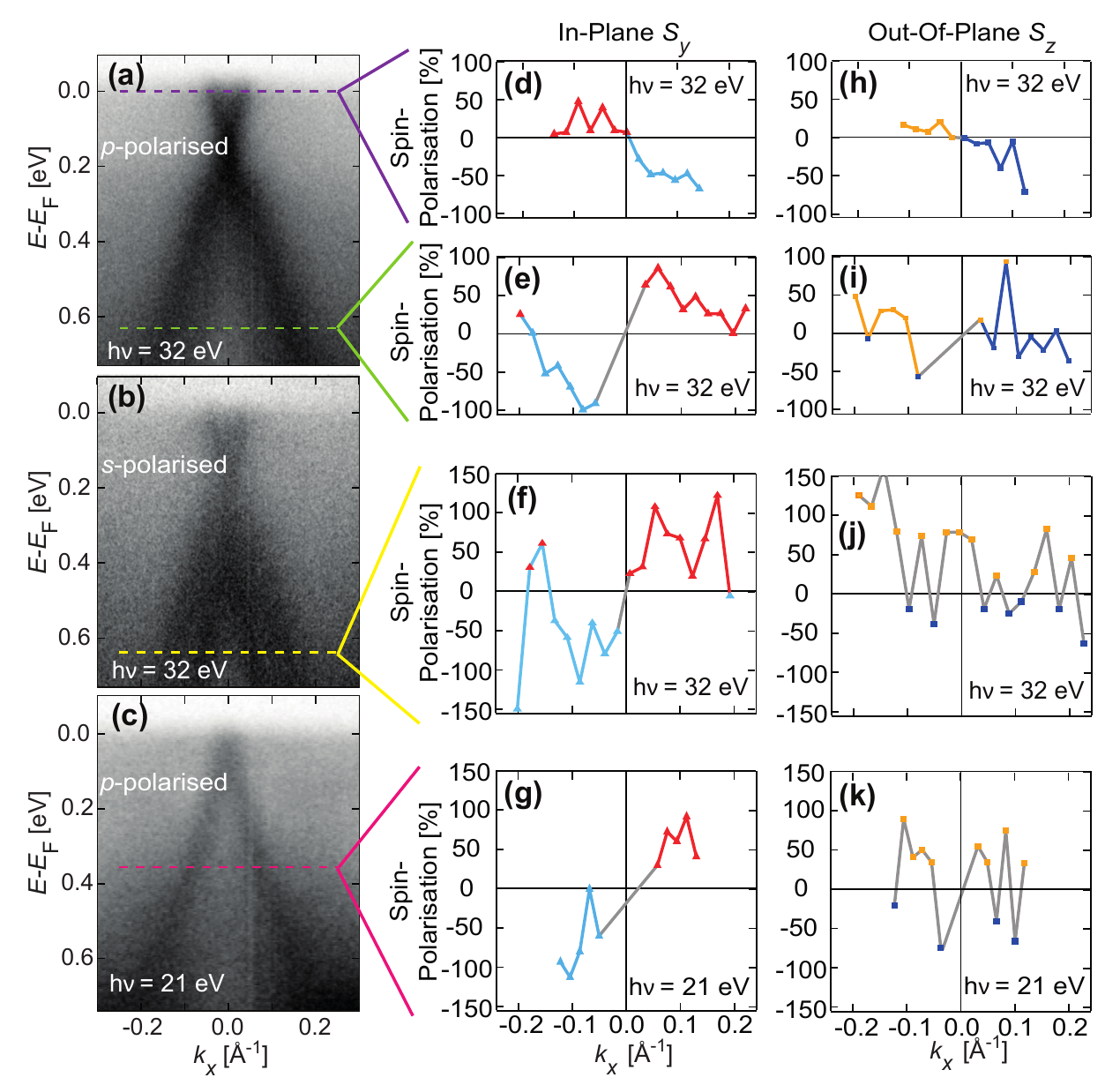}
    \caption{\textbf{(a)}-\textbf{(c)} Band structure measured with photoexcitation energy $h\nu=32$\,eV and $p$-polarised \textbf{(a)} and $s$-polarised \textbf{(b)} light, and with $h\nu=21$\,eV and $p$-polarised \textbf{(c)} light. The coloured dashed lines indicate at which binding energies ($E_{\text{B}}$) the spin-polarisation was measured. \textbf{(d)}-\textbf{(g)} In-plane spin-polarisation along the 1D line ($\vb{S}_{\text{y}}$). \textbf{(h)}-\textbf{(k)} Out-of-plane spin-polarisation ($\vb{S}_{\text{z}}$). The spin-polarisations are measured at $E_{\text{B}}=0$\,eV for \textbf{(d)} and \textbf{(h)}, at $E_{\text{B}}=0.63$\,eV for \textbf{(e)}, \textbf{(f)}, \textbf{(i)} and \textbf{(j)}, and at $E_{\text{B}}=0.36$\,eV for \textbf{(g)} and \textbf{(k)}.}
    \label{fig:SpinSuppl}
\end{figure*}

The measured spin-polarisation in-plane along the 1D lines ($\vb{S}_{\text{y}}$) and out-of-plane ($\vb{S}_{\text{z}}$) is shown at different photoexcitation energies and light-polarisations in Fig. \ref{fig:SpinSuppl}. 
The dashed lines in Figs. \ref{fig:SpinSuppl}(a)-(c) show at which binding energies the spin measurements were done. Figs. \ref{fig:SpinSuppl}(d), (e) and (g) are the same results as Figs. 4(d), (e) and (g) in the main text, respectively, but they are included to make it easier to compare with the rest of the results. 
From the main text it was seen that $\vb{S}_{\text{y}}$ reverses at the crossing points such that the spin-polarisation is positive at negative values of $\vb{k}_{\text{x}}$ at $E_\text{B}$ above the crossing point (see Fig. \ref{fig:SpinSuppl}(d)), and
positive at positive values of $\vb{k}_{\text{x}}$ below the crossing point (see Figs. \ref{fig:SpinSuppl}(e)-(g)). 
Choosing data-points only at $\vb{k}_{\text{x}}$-values where there are bands, the average $\vb{S}_{\text{y}}$ was calculated for each of the measurements. $\vb{S}_{\text{y}}=(29\pm7)\,\%$, $\vb{S}_{\text{y}}=(34\pm9)\,\%$, $\vb{S}_{\text{y}}=(55\pm 15)\,\%$ and $\vb{S}_{\text{y}}=(65\pm 11)\,\%$ for \ref{fig:SpinSuppl}(d), (e), (f) and (g), respectively.   
It is not possible to distinguish the spin-polarisation of the additional bulk bands inside $\times_{0.16\,\text{eV}}$  in Fig. \ref{fig:SpinSuppl}(b) from $\times_{0.16\,\text{eV}}$ itself, so the spin-polarisation in \ref{fig:SpinSuppl}(f) is a combination of both bulk bands and $\times_{0.16\,\text{eV}}$-electrons with $p_x$-orbital origin. 
 
Looking at $\vb{S}_{\text{z}}$ in Figs. \ref{fig:SpinSuppl}(h)-(k), the spin-reversal is not as clear as for $\vb{S}_{\text{y}}$. Figs. \ref{fig:SpinSuppl}(h) and (i) show indications of a small $\vb{S}_{\text{z}}$ component, giving average values of $\vb{S}_{\text{z}}=(12\pm4)\,\%$ and $\vb{S}_{\text{z}}=(21\pm6)\,\%$, respectively, in the region where there are bands. Fig. \ref{fig:SpinSuppl}(i) is noisier than Fig. \ref{fig:SpinSuppl}(h), probably because it is measured at an $E_{\text{B}}$ where there may be bulk bands. 
For Figs. \ref{fig:SpinSuppl}(j) and (k), $\vb{S}_{\text{z}}$ changes between positive and negative spin without a clear pattern. This indicates that there is no spin-reversal or $\vb{S}_{\text{z}}$ component for $\times_{E_\text{F}}$, and for the bands with $p_{x}$-orbital origin seen when measuring with $s$-polarised light (Fig. \ref{fig:SpinSuppl}(b)). This means that the spin-vector of $\times_{E_\text{F}}$ points in the direction of $\pm \vb{k}_{\text{y}}$. 

From the $\vb{S}_{\text{y}}$- and $\vb{S}_{\text{z}}$-components of the spin in Figs. \ref{fig:SpinSuppl}(d), (e), (h) and (i), the full spin-vector was calculated along the two dashed lines in Fig. \ref{fig:SpinSuppl}(a). The spin-vector was found to make an angle of $\theta_{\text{A}}=(22 \pm 8)^{\circ}$ at $E_{\text{B}}=0$\,eV and $\theta_{\text{B}}=(32 \pm 10)^{\circ}$ at $E_{\text{B}}=0.63$\,eV. 
Combining these, gives a spin vector for $\times_{0.16\,\text{eV}}$ which points mostly towards $\pm \vb{k}_{\text{y}}$, but makes and angle $\theta=(27\pm7)^{\circ}$ with the surface plane .  

\bibliography{Bismuth}